\documentclass{eas}
\usepackage{graphicx}
\usepackage{MR_eas}
\usepackage{natbib}
\begin{document}


\title{Two-dimensional models of early-type fast rotating stars: new
challenges in stellar physics}

\runningtitle{Two-dimensional models of early-type fast rotating stars}

\author{Michel Rieutord and Francisco Espinosa Lara}

\address{Universit\'e de Toulouse; UPS-OMP; IRAP; Toulouse, France
and CNRS; IRAP; 14, avenue Edouard Belin, F-31400 Toulouse, France}
\begin{abstract}
Two-dimensional models of rapidly rotating stars are already unavoidable
for the interpretation of interferometric or asteroseismic data of this
kind of stars.  When combined with time evolution, they will allow the
including of a more accurate physics for the computation of element
transport and the determination of surface abundances. In addition,
modeling the evolution of rotation will improve gyrochronology.

Presently, two-dimensional ESTER models predict the structure and the
large-scale flows (differential rotation and meridional circulation)
of stars with mass larger than 1.7~\msun\ at any rotation rate. Main
sequence evolution can be mimicked by varying the hydrogen content of
the convective core. Models have been successfully tested on half a
dozen of nearby fast rotating stars observed with optical or infra-red
interferometers. They are now the right tool to investigate the oscillation
spectrum of early-type fast rotators.

\end{abstract}
\maketitle
\section{Introduction}

One of the main challenges of nowadays stellar physics is the building
of two-dimensional stellar models including mean flows that pervade the
stars. These flows might have various origins but rotation is certainly
the most common.

It has long been known that rotation breaks the spherical symmetry of
a star but the contained angular momentum has also many consequences
on the internal dynamics of such stars. The Coriolis acceleration for
instance results from the conservation of angular momentum by fluid
parcels. It imposes a complicated dynamics with many types of shear
layers (Ekman or Stewartson layers) and allow new wave systems (inertial
waves). Radiative zones cannot be at rest as they are submitted to the
baroclinic torque. The resulting baroclinic flows together with spin-down
or spin-up flows coming from mass-loss or mass contraction
are at the origin of the so-called rotational mixing.
Rotational mixing designates the effects of both the meridional
circulation and the small-scale turbulence induced by differential
rotation. Much work has been devoted to its insertion in one-dimensional
codes \cite[see][for a review]{MM00}. Although undeniable successes have
been obtained with such an approach, many questions are still pending
when data are examined in closer details.  One of them is the limit of
validity of 1D models when fast rotation has to be considered.

In this contribution, we would like to show that the future of stellar
modeling is in multi-dimensional models
and, especially, in two-dimensional models. We show below that
two-dimensional models are now realistic enough to help for the
interpretation of observational data and to offer new insights in the
internal dynamics of rotating stars. A discussion of the future steps in
the improvements of such models ends this work.

\section{Why two-dimensional models?}

We may wonder why we should direct the modelling of stars towards 2D
(or 3D) models and not try to still improve 1D models that have proved so
effective in the past. The main reason comes from the strengthening
necessity of accounting for large-scale fluid flows in stars so as to
better understand the effects of secular transport on chemical
abundances. Except for a radial expansion or contraction, fluid flows
need at least two-dimensions of space to ensure mass-conservation in a
meridional motion. Indeed, using spheroidal coordinates, a steady flow
requires that

\[ \partial_\zeta(\sqrt{|g|}\rho V^\zeta) +
\partial_\theta(\sqrt{|g|}\rho V^\theta) = 0\]
with obvious notations \cite[see][]{REL13}. The meridional components of
the velocity field can therefore be expressed with a stream function
$\psi$ such that

\[ V^\zeta = (\partial_\theta\psi)/\rho\sqrt{|g|}, \qquad V^\theta =
-(\partial_\zeta\psi)/\rho\sqrt{|g|}\]
It shows that vertical transport (by $V^\zeta$) requires latitudinal
variations of the stream function. Hence, 2D models are the minimum step
to include fluid flows in a somewhat realistic way.

However, beyond this a priori theoretical requirement, observational
data also force us to forget about the spherical symmetry. Let us
discuss the main points.

\subsection{Images of fast rotators}

\begin{figure}[h]
\begin{minipage}[t]{0.47\linewidth}
\centerline{\includegraphics[width=\linewidth,angle=0]{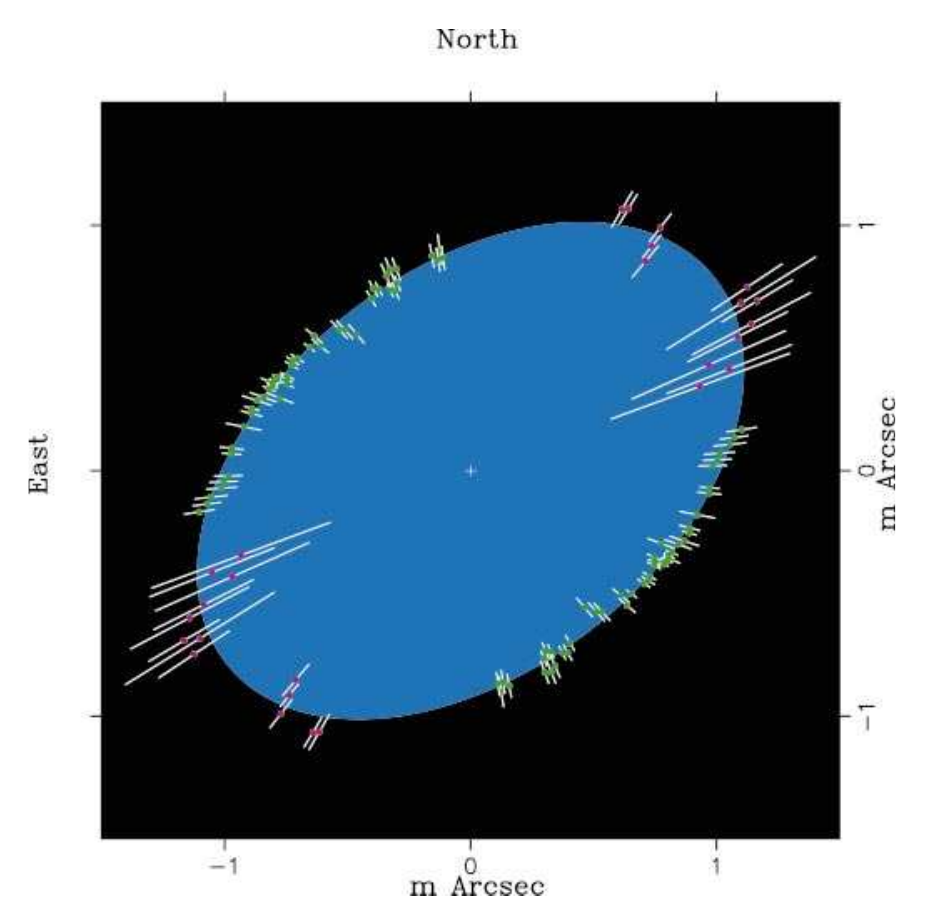}}
\caption[]{First determination of Achernar's shape with VLTI
\cite[][]{domiciano_etal03}.}
\label{image_Ach}
\end{minipage}
\hfill
   \begin{minipage}[t]{0.47\linewidth}
\includegraphics[width=\linewidth,angle=0]{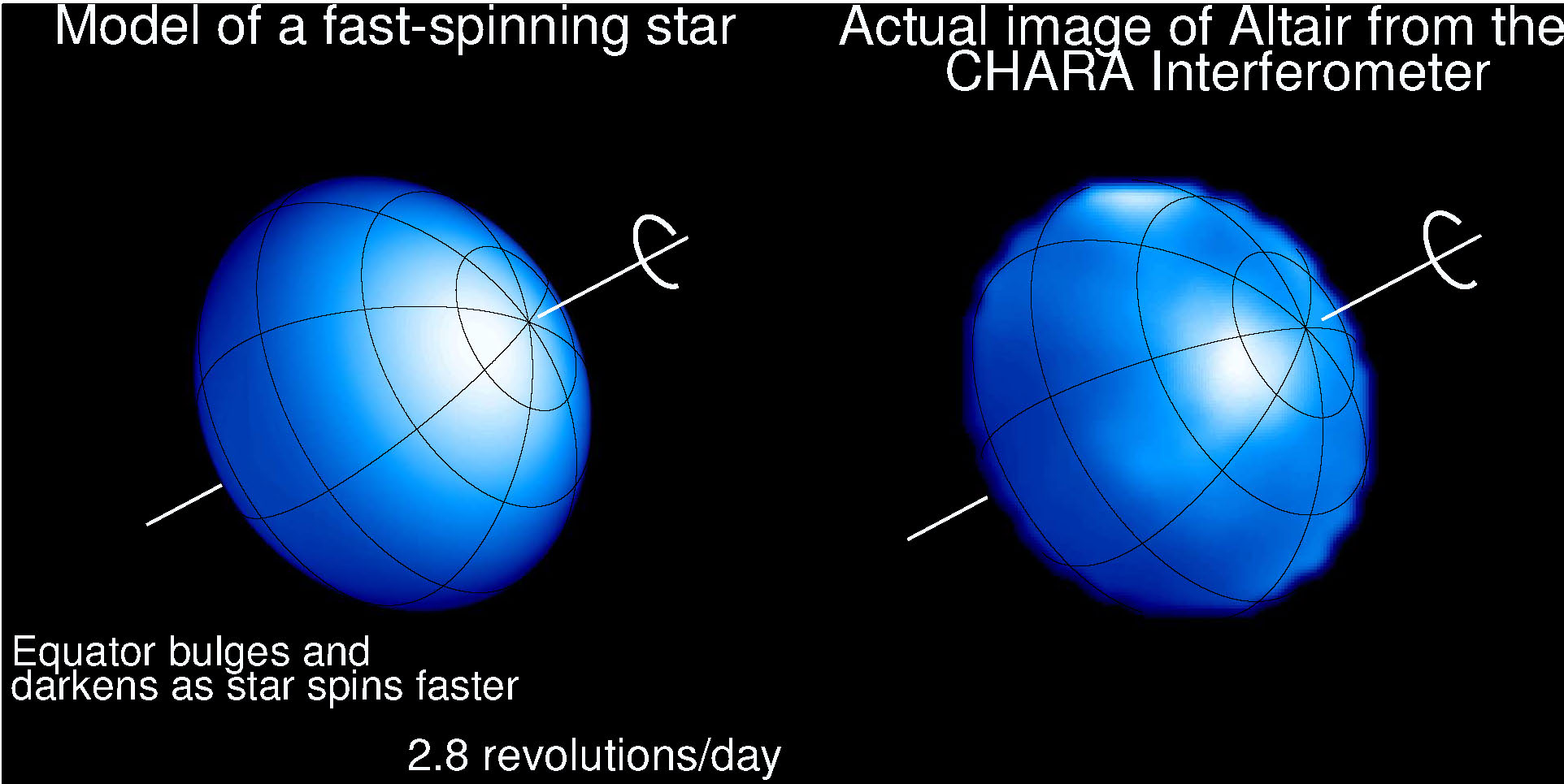}
\caption[]{Reconstruction of the surface brightness of Altair with CHARA
interferometer by \cite{monnier_etal07}.}
\label{image1}
\end{minipage}
\end{figure}

Since the work of \cite{domiciano_etal03} on Achernar (see
Fig.~\ref{image_Ach}), imaging the surface of nearby rotating stars
with optical or IR interferometers has made substantial progresses
\cite[e.g.][and Fig.~\ref{image1}]{monnier_etal07}. Presently, the shape
and the basic brightness distribution is measured for a ten of stars. But
the derivation of fundamental parameters of these stars relies on
models that are used to match interferometric visibilities or closure
phases. So the better the models, the better the parameters. Up to now
\cite[e.g.][]{monnier_etal12}, Roche models combined with von Zeipel
laws of gravity darkening are used to invert these data.  These models
are very basic and more realistic ones are extremely welcome to give
more accurate measurements.

\begin{figure}[h]
\centerline{\includegraphics[width=0.8\linewidth,angle=0]{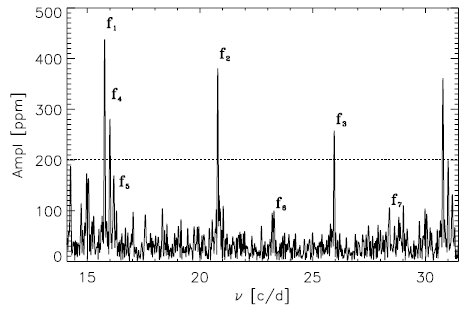}}
\caption[]{Part of the oscillation spectrum of Altair from WIRE
\cite[][]{buzasietal05}.}
\label{altairfreq}
\end{figure}

\subsection{Asteroseismology}

The large amount of photometric data that have been collected by space missions
like CoRoT, KEPLER, MOST or WIRE include a large number of rapidly
rotating stars.  These data, like those shown in Fig.~\ref{altairfreq},
are waiting for an interpretation. This interpretation is of crucial
importance to get a direct view on the interior of early-type stars. Apart
a few exception, all these stars rotate rapidly. They are powered by the
CNO cycle in a convective core.  A long-standing question about them is
the size of this core and the amount of overshooting of convection in the
envelope. As shown by first investigations \cite[e.g.][]{R06,ELR13}, the
core-envelope interface is likely a complicated region from the dynamical
point of view. Asteroseismic constraints are therefore most welcome to
select the right models.  Since fast rotation is an unavoidable feature
of these stars, 2D models are necessary to identify, interpret
and invert the associated asteroseismic data. The CoRoT target HD43317,
a main sequence B-type star rotating at 50\% of its critical velocity, is
certainly the best illustration of this need \cite[see][]{papics_etal12}.

\subsection{Abundance patterns}

1D-models have long been used to investigate the surface abundances in
relation with rotational mixing \cite[e.g.][]{maeder09} with true
success. However, recent works based on large spectroscopic surveys with
the VLT, comparing observed abundance of nitrogen and prediction of
models with the so-called Hunter diagram \cite[][]{brott_etal11} show
obvious discrepancies between models an observations. The long-standing
questions about the abundances of lithium or beryllium are still
stimulating research on the way these elements are transported and
depleted \cite[e.g.][]{tognelli_etal12}. Clearly, more refined
models that describe mixing associated with rotation are needed.

\subsection{Other fundamental questions about rotating stars}

Even if observations prompt us to difficult questions (like the previous
ones), the conception of models or simply our curiosity
about the evolution of stars, rises other fundamental questions on the
role of rotation in stellar evolution. A few of them are the following:

\begin{itemize}
\item What is the dynamical evolution of rotating PMS stars? In which
dynamical state do they reach the main sequence?

\item First stars, so-called population III stars, were lacking metals
and therefore were more compact, concentrating angular momentum. What
has been the evolution and the yields of these stars whose rotation
has been certainly quite fast?

\item It is conjectured that rotation plays a crucial role in the
explosion of supernovae that produce gamma ray burst. What is the
pre-supernova dynamical state?

\end{itemize}

\subsection{Conclusions}

All the foregoing arguments show that two-dimensional models including
rotation at any rate will permit the detailed exploration of numerous
very timely problems. It is one of the grand challenges of stellar
physics for the next decades.

\section{Constructing 2D models}

\subsection{Basic assumptions}

In order to make the problem well-posed and tractable, we consider an
isolated, non-magnetic rotating star. Presently, we discard any mass-loss or
time-evolution and search for steady state solutions. We thus look
for the structure of such stars including their large-scale flows. Such
models may not be very far from actual early-type stars
(typically A and B stars).

\subsection{Equations to be solved}

Let us just recall that the foregoing problem is that of the flow of a
self-gravitating compressible plasma. It is controlled by the four
following partial differential equations:

\greq
     \Delta\phi = 4\pi G\rho \\
     \rho T \vv\cdot\na S = -\Div\vF + \eps_*\\
     \rho (2\vO_*\wedge\vv + \vv\cdot\na\vv) = -\na P
    -\rho\na(\phi-\textstyle{\demi}\Omega_*^2s^2)+\vF_v\\
     \Div(\rho\vv) = 0.
\egreqn{basiceq}
namely Poisson equation, entropy equation, momentum equation and mass
conservation equation respectively.  These equations are completed by
the microphysics prescriptions

\greq
     P\equiv P(\rho,T) \quad {\rm OPAL}\\
     \khi_r \equiv \khi_r(\rho,T) \quad {\rm OPAL}\\
     \eps_* \equiv \eps_*(\rho,T)\quad {\rm NACRE}
\egreq
and some transport prescriptions:

\begin{itemize}
\item The energy flux
\[
\vF = -\khi_r\na T -\frac{\khi_{\rm turb}T}{{\cal R}_M}\na S
\]

\item The transport of momentum
\beqa
 \vF_v = \mu\vec{\cal F}_\mu(\vv) &=&
\displaystyle\mu\lc\Delta\vv+\frac{1}{3}\na\left(\na\cdot\vv\right)
+2\left(\na\ln\mu\cdot\na\right)\vv\right.\nonumber \\
&&\displaystyle
\quad\left.+\na\ln\mu\times(\na\times\vv)
-\frac{2}{3}\left(\na\cdot\vv\right)\na\ln\mu\rc \;.
\eeqa
or any mean-field expression of the Reynolds stress.

\end{itemize}

\noindent Boundary conditions read

\begin{itemize}
\item On pressure (radiative equilibrium)
\[ P_s=\frac{2}{3}\frac{\overline{g}}{\overline{\kappa}}\]
\item On the velocity field (stress-free conditions)
\[ \vv\cdot \vn =0 \quad {\rm and}\quad ([\sigma]\vn)\wedge\vn =\vzero
\]
\item On temperature (black body radiation)
\[ \vn\cdot\na T + T/L_T=0 \]
\end{itemize}
with a prescription of the angular momentum content or of the equatorial
velocity, namely:

\[ \intvol r\sth\rho u_\varphi \,dV = L \qquad {\rm or}\qquad
 v_\varphi(r=R,\theta=\pi/2) = V_{\rm Eq} \]
Detailed comments may be found in \cite{ELR13}.

\begin{figure}[h]
\centerline{
\includegraphics[width=0.8\textwidth,angle=0]{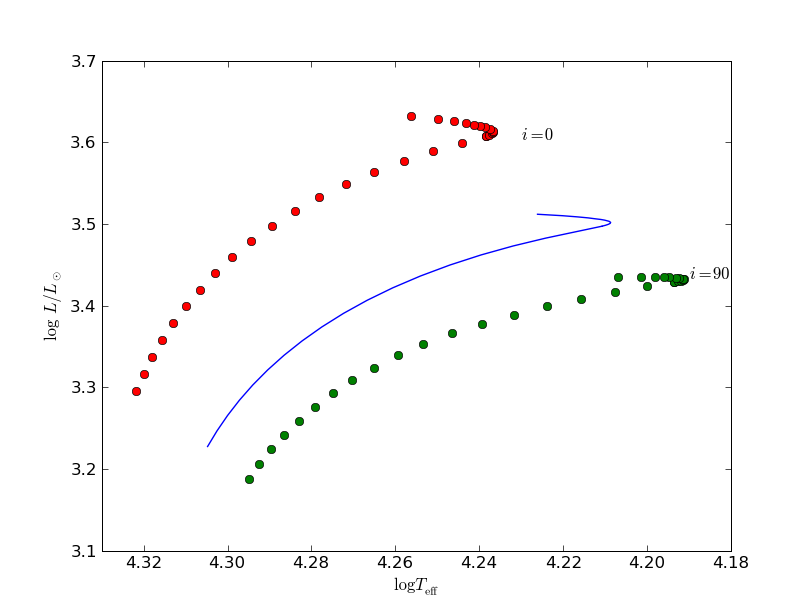}}
\centerline{
\includegraphics[width=0.8\textwidth,angle=0]{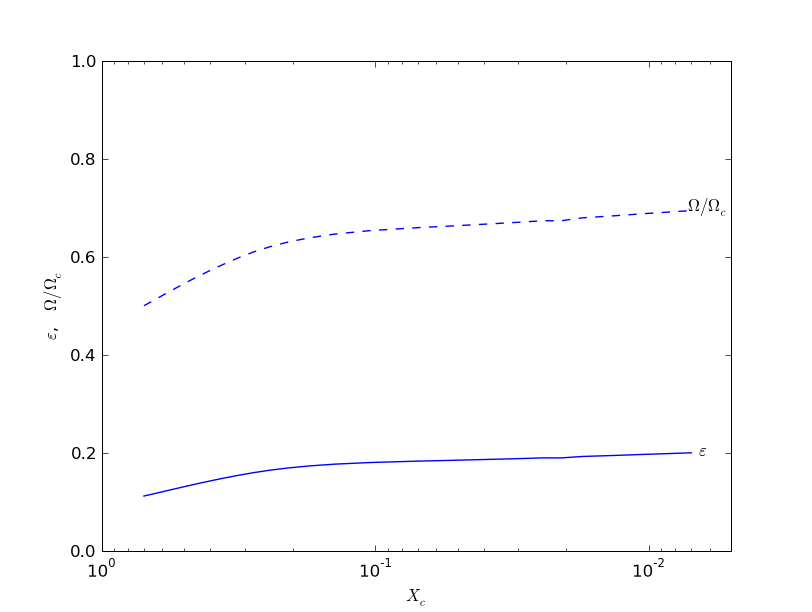}
}
\caption[]{Variation of the position of a 7\msun \  star in the HR diagram
when the hydrogen
content of its core is decreased. The apparent trajectory that depends
on the inclination $i$ of the rotation axis with respect to the line of
sight is given for $i=0$ (pole on) and $i=90^\circ$ (equator on).}
\label{HR}
\end{figure}

\section{The ESTER project and some results}

\subsection{Present status}

We may briefly present the ESTER code as follows:

\begin{itemize}
\item It is a spectral code (Chebyshev polynomials \& spherical harmonics)
written in C++ \& Python. It is an open source code
(http://code.google.com/p/ester-project).
\item The spheroidal shape of  the star is accounted through a mapping of the
natural coordinates towards spherical ones \cite[see][]{REL13}.
\item It solves the structure equations in the asymptotic limit of small
viscosities with the Newton algorithm.
\item It has been tested through comparisons with 1D codes at $\Omega=0$.
\item Solutions are checked with internal tests (virial, energy integral,
spectral convergence, roundoff errors).

\end{itemize}

Presently, the ESTER code can compute models for early-type stars of mass
larger than 1.7~\msun\ rotating up to breakup. In Tab.~\ref{alder}, we give
the example of Alderamin ($\alpha$ Cep) and Altair ($\alpha$ Aql)
showing that our models nicely
fit the observed fundamental parameters of these stars as determined by
interferometry. This is also the case for other A-type stars like Vega
or Rasalhague or the B-star $\alpha$ Leo \cite[see][]{ELR13}. Three other stars,
$\delta$ Vel Aa, Ab and Achernar ($\alpha$ Eri) have also been investigated in
\cite{REL13b}. In these models, there
is no time evolution but chemical evolution on the main sequence can
be mimicked by changing the hydrogen content of the core. In this way
we have explored the evolution along the main sequence for stars of
constant angular momentum.

\begin{table}
\caption{Comparison between observationally derived
parameters of the stars and tentative two-dimensional models. Data from
Altair are from \cite{monnier_etal07} and those of Alderamin are
from \cite{zhao_etal09}. Here $\eps$ is the flatness, $\omega_k$ the
ratio of the equatorial angular velocity to the keplerian one, $j$ is
the mean specific angular momentum of the star and X$_\mathrm{env.}$ is
the hydrogen mass fraction in the envelope.}
\vspace*{10pt}
\begin{center}
\begin{tabular}{lllll}
\hline
Star                     & \multicolumn{2}{c}{Altair ($\alpha$ Aql}) &
\multicolumn{2}{c}{Alderamin ($\alpha$ Cep}) \\
                         & Observations& Model  & Observations& Model \\
                         &          &        &          &       \\
Spectral type            & A7 IV-V  &        & A8 V    &       \\
Mass  (M$_\odot$)        &               & 1.65 & 1.92$\pm0.04$ & 1.92 \\
R$_{\rm eq}$ (R$_\odot$) & 2.029$\pm$0.007 & 2.027 & 2.739$\pm$0.04 & 2.747 \\
R$_{\rm pol}$ (R$_\odot$)& 1.634$\pm$0.011 & 1.627& 2.199$\pm$0.035& 2.166 \\
$\eps$      &         & 0.197&           & 0.211  \\
$\omega_k$  &         & 0.695&           & 0.725 \\
$i$         &     57$^\circ$ &    & 56$^\circ$&   \\
T$_{\rm eq}$ (K)         & 6860$\pm$150  & 6849  &6574$\pm$200   & 6781 \\
T$_{\rm pol}$(K)         & 8450$\pm$140  & 8450  &8588$\pm$300   & 8515 \\
L (L$_\odot$)            &               & 10.2 & 18.1$\pm$1.8   & 18.6 \\
V$_{\rm eq}$ (km/s)      & 285.5$\pm$6   & 274  & 272$\pm$18  & 265 \\
j (10$^{17}$ cm$^2$/s)    &               & 1.12 &            & 1.27 \\
k [j/R$_{\rm eq}^2\Omega_{\rm eq}$] &    & 0.0289 &           & 0.0251 \\
P$_{\rm eq}$ (days)      &               & 0.374&             & 0.525 \\
P$_{\rm pol}$ (days)     &               & 0.387&             & 0.538 \\
X$_\mathrm{env.}$        &               & 0.70 &           & 0.70 \\
X$_\mathrm{core}$/X$_\mathrm{env.}$&     & 0.50 &           & 0.30 \\
Z                        &               & 0.014&           & 0.017\\
\hline
\end{tabular}
\end{center}
\label{alder}
\end{table}

The results of such calculations are shown in Fig.~\ref{HR}. In this case
the initial rotation rate is 50\% critical. We note that changing the
orientation of the rotation axis may change the apparent luminosity by up to
a factor 2 (note that limb darkening is not accounted for in this
calculation of the apparent luminosity). This same figure is completed
by the ``evolution" of the angular velocity with respect to critical and
of the flatness of the surface. As obvious in this diagram,  evolution
on the main sequence leads to a more and more critical rotation of the
star. This comes from the fact that even if the equatorial velocity
diminishes because of angular momentum conservation and the growth of
the radius, the critical equatorial velocity decreases even more rapidly,
making the star's rotation closer and closer to the critical one.

This result clearly shows that stars born at sufficiently rapid rotation
rate (typically 60\% critical) will reach the critical rotation on the
main sequence. It supports the idea that the Be phenomenon is the result
of the evolution of initially fast rotating B-stars, although
observations do not seem to show unambiguously a frequency rise of Be
stars at the end of the main sequence \cite[][]{PR03}.

\section{Conclusion}

Presently, the ESTER code can compute rather easily models of isolated
interme\-diate-mass or massive stars at any rotation rate. The
models are self-consistent in the sense that they do not depend on
an arbitrary differential rotation. Their differential rotation (and
associated meridional circulation) are computed as the result of the
baroclinic torque existing in the radiative envelope. The comparisons
of fundamental parameters of the few nearby fast rotators observed with
interferometry with ESTER models are quite satisfactory. These models
can be used to invert interferometric visibilities and closure phases
as they give more realistic gravity darkening profiles than Roche/von
Zeipel models \cite[][]{ELR11}. The next step in comparing models and
data is that of the identification of eigenmodes in a fast rotating star
\cite[][]{mirouh_etal13,reese_etal13b}, which is still a challenge.

The future of 2D stellar models is now to move from steady state models to
time evolving ones with either gravitational contraction or
nuclear reactions networks. As far as structure is concerned, we also plan to
extend the capabilities of the code to low-mass stars, but this needs
the development of an efficient modeling of convective envelopes in
two-dimensions.

\bibliographystyle{aa}
\bibliography{/home/rieutord/tex/biblio/bibnew}

\end{document}